\documentclass[superscriptaddress,twocolumn]{revtex4}
\usepackage{amsmath,lscape,epsfig}

\def\beq{\begin{equation}}
\def\eeq{\end{equation}}
\def\beqa{\begin{eqnarray}}
\def\eeqa{\end{eqnarray}}
\def\ban{\begin{eqnarray*}}
\def\ean{\end{eqnarray*}}
\def\bi{\begin{itemize}}
\def\ei{\end{itemize}}

\begin{document}

\title{Reexamining the neutron skin thickness within a density
  dependent hadronic model}

\author{S.S. Avancini}
\affiliation{Depto de F\'{\i}sica - CFM - Universidade Federal de Santa
Catarina  Florian\'opolis - SC - CP. 476 - CEP 88.040 - 900 - Brazil}
\author{J.R. Marinelli}
\affiliation{Depto de F\'{\i}sica - CFM - Universidade Federal de Santa
Catarina  Florian\'opolis - SC - CP. 476 - CEP 88.040 - 900 - Brazil}
\author{D.P. Menezes}
\affiliation{Depto de F\'{\i}sica - CFM - Universidade Federal de Santa
Catarina  Florian\'opolis - SC - CP. 476 - CEP 88.040 - 900 - Brazil}
\author{M.M.W. Moraes}
\affiliation{Depto de F\'{\i}sica - CFM - Universidade Federal de Santa
Catarina  Florian\'opolis - SC - CP. 476 - CEP 88.040 - 900 - Brazil}
\author{A.S. Schneider}
\affiliation{Depto de F\'{\i}sica - CFM - Universidade Federal de Santa
Catarina  Florian\'opolis - SC - CP. 476 - CEP 88.040 - 900 - Brazil}

\begin{abstract}
In the present work we calculate the $^{208}$Pb neutron skin
thickness, binding energy, surface energy and density profiles within the
Dirac solution of a density dependent hadronic model. The same calculation is
performed with the NL3 parametrization of the non-linear Walecka model.
The asymmetry of a polarized electron scattered from a heavy target is also
obtained within a partial wave expansion calculation. The theoretical results
are then ready to be compared with the experimental results expected to be
available very soon at the Jefferson Laboratory. For completeness, other
nuclei as $^{40}$Ca, $^{48}$Ca,$^{66}$Ni and $^{90}$Zr are also investigated.
\end{abstract}

\maketitle

\vspace{0.50cm}
PACS number(s): {21.65.+f,24.10.Jv,95.30.Tg,26.60.+c}
\vspace{0.50cm}

\section{Introduction}

Relativistic models are a very useful tool in the description of a wide
variety of applications in nuclear matter, finite nuclei
and nuclear astrophysics. They can be
tested and, hopefully constrained, according to experimental and astrophysical
observation results. Unfortunately, most theoretical results are
model dependent and so far it is unclear whether some of them should be
discarded. Many variations of the well known
quantum hadrodynamic model \cite{sw} have been developed and used along the
last decades. Some of them rely on density dependent couplings between the
baryons and the mesons \cite{original,tw,br,gaitanos} while others use
constant couplings \cite{nl3,tm1,glen}. Still another possibility of including
density dependence on the Lagrangian density is through the coupling of the
mediator mesons among themselves \cite{nlwr}.
The strong model dependence of the results come from the simple fact that
relativistic model couplings  are adjusted in
order to fit expected nuclei properties such as binding energy, saturation
density, compressibility and energy symmetry at saturation density only.
Once the same relativistic models are extrapolated to higher
densities as in stellar matter or higher temperatures as in heavy-ion
collisions or even to lower densities as in the nuclear matter liquid gas
phase transitions, they can and indeed provide different information. Hence,
experimental constraints obtained either from polarized electron scattered
from a heavy target, from heavy-ion collisions at different energies or from
astronomical observations are very important in order that adequate models
are chosen and inadequate ones are ruled out.

In the present work we focus our attention to the calculation of
the difference between the neutron and the proton radii known as
the neutron skin thickness.
Based on the argument that experimental results should be used to constrain
relativistic models, it is very important that an
accurate experimental measurement of the neutron skin thickness is
achieved.  This depends on a precise measurement of both
the charge and the neutron radius. The charge radius is already known
within a precision of one  percent for most stable nuclei, using the
well-known single-arm and non-polarized elastic electron
scattering technique  as well as the spectroscopy of muonic atoms
\cite{vries}. For the neutron radius, our present knowledge has
an uncertainty of about 0.2 fm \cite{horo}. However, using
polarized electron beams it is possible to obtain the neutron
distribution, as first
discussed in \cite{Don} and, as a consequence, to obtain the
desired neutron radius. In fact, the Parity Radius Experiment
(PREX) at the Jefferson Laboratory \cite{prex} is currently
running to measure the $^{208}$Pb neutron radius with an accuracy
of less than 0.05 fm, using polarized electron scattering.
In the PREX experiment, the asymmetry is expected to be measured
at a momentum transfer $q\approx0.4~fm^{-1}$.

At this point it is worth mentioning that two other experimental methods have
been used in order to measure neutron skins and neutron halos
\cite{trzcinska,klos}, namely, the nuclear spectroscopy analysis of the
antiproton annihilation residues one mass unit lighter than the target mass
and the measurements of strong-interaction effects on anti-protonic X rays.
Whenever possible, the results obtained with the above mentioned methods are
also used for comparison in the results section.

In a recent work \cite{peles}, the neutron skin thickness and the asymmetry
for polarized electron scattering off a hadronic target were investigated
with the help of both density dependent and with constant coupling
relativistic models. For these calculations two important compromises were
made: for the calculation of the proton and neutron densities, a Thomas-Fermi
approach was used and for the calculation of the asymmetry a Plane Wave Born
Approximation for the electron \cite{horo98} was enforced.
The $^{208}Pb$ neutron skin thickness was then obtained with two different
  density dependent hadronic (DDH) model parametrizations, the TW \cite{tw},
where the density dependence is
  introduced explicitly through the couplings and the NL$\omega\rho$ model
\cite{nlwr}, where the density dependence appears through the coupling of the
vector and
the isovector mesons. One of the most used parametrizations of the non-linear
Walecka model (NLWM), the NL3 was also used. The asymmetry, in the momentum
transfer range of interest for the calculation of neutron skins, was shown to
give very
  similar results for all models. On the other hand, the neutron skin
thickness is smaller with the DDH model than with the NL3. As the coupling
strength between the $\omega$ and $\rho$ mesons increases in the
NL$\omega\rho$ model, the neutron skin
thickness moves from the original NL3 towards the DDH results.
As the momentum transfer increases, the asymmetry becomes strongly model
dependent.

In the present work we revisit the same quantities, but with improvements in
both approximations mentioned above in order to check if the model differences
remain.
As the target is a heavy
nucleus ($^{208}$Pb), the results for the asymmetry are reconsidered with the
use of the partial wave expansion method as briefly discussed in the
Appendix. On the other hand,
the proton and neutron densities are calculated from the solution of the Dirac
equation. We also extend our calculations to obtain the neutron
skin thickness of $^{40}Ca$, $^{48}Ca$,$^{66}Ni$ and $^{90}$Zr. For the
asymmetry, we also include applications for $^{48}Ca$.
As the NL$\omega\rho$ model was shown to give
results that interpolate between the NL3 parametrization of the NLWM and the
TW  parametrization of the DDH model, we next
restrict ourselves to the NL3 and the DDH models, i.e., one with constant
couplings and another one with density dependent couplings.
We also comment on the differences between the results obtained
in the present work and within the Thomas-Fermi approximation. It is worth
mentioning that most applications to
  neutron stars, to equations of state used to describe supernova simulations
and the description of nucleation processes have been done within the
  Thomas-Fermi approximation. Understanding its limitations and accuracy is
  indeed very important. Information about the skin thickness as obtained in
finite nuclei can be directly related to astrophysical calculations and so the
possible connections between Thomas-Fermi and Dirac density profiles can be
useful.

The paper is organized as follows: in section II we show the Lagrangian
density of the DDH model and describe the formalism used; in section
III we present and discuss the results; in section IV we draw our final
conclusions.

\section{Formalism}

We describe the main quantities of the DDH model, which has
density dependent coupling parameters in the following.
The Lagrangian density reads:
$$
{\cal L}_H=\bar \psi\left[\gamma_\mu\left(i\partial^{\mu}-\Gamma_v V^{\mu}-
\frac{\Gamma_{\rho}}{2}  \boldsymbol{\tau} \cdot \mathbf {b}^\mu \right. \right.
$$
$$ \left. \left. -e  \frac{(1+\tau_{3})}{2} A^\mu \right)
-(M-\Gamma_s \phi)\right]\psi
$$
$$
+\frac{1}{2}(\partial_{\mu}\phi\partial^{\mu}\phi
-m_s^2 \phi^2)
-\frac{1}{4}\Omega_{\mu\nu}\Omega^{\mu\nu}$$
\begin{equation}
+\frac{1}{2}m_v^2 V_{\mu}V^{\mu}
-\frac{1}{4}\mathbf B_{\mu\nu}\cdot\mathbf B^{\mu\nu}+\frac{1}{2}
m_\rho^2 \mathbf b_{\mu}\cdot \mathbf b^{\mu}
-\frac{1}{4}F_{\mu\nu}F^{\mu\nu}
\label{lagtw}
\end{equation}
where $\phi$, $V^\mu$, $\mathbf {b}^\mu$ and $A^{\mu}$ are the
scalar-isoscalar, vector-isoscalar and vector-isovector meson
fields and the photon field respectively,
$\Omega_{\mu\nu}=\partial_{\mu}V_{\nu}-\partial_{\nu}V_{\mu}$ ,
$\mathbf B_{\mu\nu}=\partial_{\mu}\mathbf b_{\nu}-\partial_{\nu} \mathbf b_{\mu}
- \Gamma_\rho (\mathbf b_\mu \times \mathbf b_\nu)$,
$F_{\mu\nu}=\partial_{\mu}A_{\nu}-\partial_{\nu}A_{\mu}$
and $\tau_{3}=\pm 1$ for protons and neutrons respectively. The  parameters of
the model are:
the nucleon mass $M=939$ MeV, the masses of the mesons $m_s$, $m_v$,
$m_\rho$, the electromagnetic coupling constant $e=\sqrt{4\pi/137}$
and the density dependent couplings $\Gamma_{s}$,
$\Gamma_v$ and $\Gamma_{\rho}$, which are adjusted in order to reproduce
some of the nuclear matter bulk properties,
using the following parametrization:
\begin{equation}
\Gamma _{i}(\rho )=\Gamma _{i}(\rho _{sat})h_{i}(x),\quad x=\rho /\rho _{sat},
\label{paratw1}
\end{equation}
with
\begin{equation}
h_{i}(x)=a_{i}\frac{1+b_{i}(x+d_{i})^{2}}{1+c_{i}(x+d_{i})^{2}},\quad i=s,v
\end{equation}
and
\begin{equation}
h_{\rho }(x)=\exp [-a_{\rho }(x-1)],  \label{paratw2}
\end{equation}
with the values of the parameters $m_{i}$,
$\Gamma _{i}(\rho_{sat})$, $a_{i}$, $b_{i}$, $c_{i}$ and $d_{i}$,
$i=s,v,\rho $ given in \cite{tw}.
This model does not include self-interaction terms for the meson
fields as in NL3. We consider two parametrizations of the above mentioned
DDH model, the original one that we next refer to as TW and a more recent one
known as DDME1 \cite{twring}, obtained from a fitting that includes known
experimental values of the $^{208}Pb$ neutron skin.

Once the Lagrangian density is chosen, the Euler-Lagrange equations are used to
calculate the equations of motion. The meson field equations of motion are
easily found in the literature and we refrain from writing them. An interested
reader can obtain the equations in \cite{original,tw,ddpeos,peles}, among
other papers in the literature.

The Dirac equation for the nucleon field reads:
\begin{equation} \label{eq1}
[ \gamma^\mu (i \partial_\mu - \Sigma_\mu) - (M- \Sigma_s) ] \Psi = 0 ~~,
\end{equation}
where the scalar and vector self-energies are given respectively by

\begin{equation} \label{escalar}
\Sigma_s = \Gamma_s \phi, \quad
\Sigma_\mu =\Sigma_\mu^{(0)}  +\Sigma_\mu^R,
\end{equation}
with
\begin{equation}
\label{vetorial}
\Sigma_\mu^{(0)} = \Gamma_v V_\mu +  \frac{\Gamma_\rho}{2}
\vec{\tau} \cdot {\mathbf b}_\mu + e\frac{(1+ \tau_3)}{2}
A_\mu ~~,
\end{equation}
\[ \Sigma_\mu^R = \frac{j_\mu}{\rho} \left( \frac{\partial \Gamma_v}{\partial
    \rho} \bar{\psi} \gamma^\nu \psi V_\nu + \frac{1}{2} \frac{\partial
    \Gamma_\rho}{\partial \rho} \bar{\psi} \gamma^\nu \vec{\tau} \psi \cdot
  {\mathbf b}_\nu
- \frac{\partial\Gamma_s}{\partial \rho} \bar{\psi}\psi \phi \right). \]
The term $ \Sigma_\mu^R $ is known as the rearrangement term.
As done in \cite{ring,sw}, we consider that the nucleus is approximately given
by a Slater determinant and only the lowest lying positive energy states are occupied. The
usual ansatz for the nucleon spinor is
\begin{equation} \label{ansa}
\psi ( \vec{{\rm{r}}},t ) =\psi(\vec{{\rm{r}}}) \exp (-iE t),
\end{equation}
and the one-particle states are obtained from the solution of the Dirac
equation. Only spherically symmetric nuclei are considered and the usual
notation for the expected values of the meson fields are used. The
self-energies become:
\begin{equation} \label{sigma1}
\Sigma_0 = \Gamma_v V_0(\rm{r}) + \frac{1}{2} \Gamma_\rho \tau_3
b_0(\rm{r}) + e\frac{(1+ \tau_3)}{2} A_{0}(\rm{r}) + \Sigma_0^R ~~,
\end{equation}
\begin{equation} \label{sigma2}
 \Sigma_0^R  = \left( \frac{\partial \Gamma_v}{\partial \rho} \rho(r) V_0
  + \frac{1}{2}\frac{\partial \Gamma_\rho}{\partial \rho} \rho_3(r) b_0  -
  \frac{\partial\Gamma_s}{\partial \rho} \rho_s(r) \phi \right),
\end{equation}
where $\rho_3=\rho_p-\rho_n$ and $\rho_s$ and $\rho$ are the usual scalar and baryonic densities.
Next the ansatz given in eq.(\ref{ansa}) and the spin-isospin wavefunctions
$ \psi = \left( \begin{array}{c} g_\kappa(r) \mathcal{Y}_\kappa^{jm}
\\ i f_\kappa(r) \mathcal{Y}_{-\kappa}^{jm} \end{array} \right)\otimes
\xi$
are substituted into the Dirac equation.
$\mathcal{Y}_{\pm \kappa}^{jm}$ and $\xi$ are
the spinorial spherical harmonics and the isospin wavefunctions respectively.
After some straightforward algebraic
manipulations, a new system of coupled equations is obtained:
\[ \left(M^\star(r) + V(r)  \right) g_\kappa(r) - \left( \frac{
    \partial}{\partial r} - \frac{\kappa-1}{r} \right) f_\kappa(r) = E
g_\kappa(r) \]
\begin{equation}
\label{sis2} \left( \frac{ \partial}{\partial r} + \frac{\kappa+1}{r} \right)
g_\kappa(r) - \left(M^\star(r) - V(r) \right) f_\kappa(r) = E f_\kappa(r) ~~,
\end{equation}
where
$$
M^{\star}(r)=M-\Gamma_s(r) \phi(r),
$$
and
$$
V(r)=\Gamma_v(r) V_0(r)+ \frac{\Gamma_\rho(r)}{2} \tau_3 b_0(r) +
e\frac{(1+ \tau_3)}{2} A_{0}(r) +\Sigma_0^R(r).
$$
At this point, $f_{\kappa}(r)$ and $g_{\kappa}(r)$ are expanded in the
harmonic oscillator basis of dimensions $N$ and $M$ respectively, as in \cite{ring}.
The linear variational method is then used to solve the above equations and
we are left with an eigenvalue problem solved by the diagonalization of a
matrix of order $N+M$. The fields $\phi$, $V_0$ and $b_0$ are also expanded in
a harmonic oscillator basis and solved iteratively through their corresponding
Klein-Gordon equations. The Coulomb field $A_0$ is obtained by the Green's
function method.

The proton and neutron mean-square radius are defined as

\begin{equation}
R_{i}^{2}=\frac{\int~d^{3}r r^2 \rho_{i}(\mathbf {r})}{\int~d^{3}r
 \rho_{i}(\mathbf {r})}, \quad i=p,n
\end{equation}
\noindent and the neutron skin thickness reads
\begin{equation}
\theta=R_n-R_p. \label{skin}
\end{equation}

It is also useful to define the nuclear charge distribution as:
\begin{equation}
\rho_c(\overrightarrow{r})=\frac{1}{[(a_p^{2}-B^{2})\pi]^{3/2}}\int~d^{3}r'exp[-\frac{(\overrightarrow{r}-
\overrightarrow{r}')^2}{(a_p^{2}-B^{2})}]\rho_p(\overrightarrow{r}'), 
\label{rhoc}
\end{equation}

\noindent where $B^2=b^2/A$, $a_p= 0.653$ fm and $b$ is the oscillator length
used to define our basis set. The above definition includes the proton finite 
size, for which we have assumed a gaussian shape to describe its intrinsic 
charge distribution:
 
 \begin{equation}
 \rho_{int}(r)=\frac{1}{(a_{p}^{2}\pi)^{3/2}}~exp~(-(r/a_p)^2),
 \end{equation}
 
 \noindent and an approximate center-of-mass correction according to the 
prescription given in \cite{negele}.
 Neutron finite size corrections are disregarded, i.e., its intrinsic form 
factor is taken as unit.

For the calculation of the differential elastic electron cross-section
($\frac{d\sigma}{d\Omega}$) and the asymmetry ($\cal{A}$), details are given
in the Appendix.

\section{Results and discussion}

\begin{figure}[b]
\begin{center}
\includegraphics[width=8.cm]{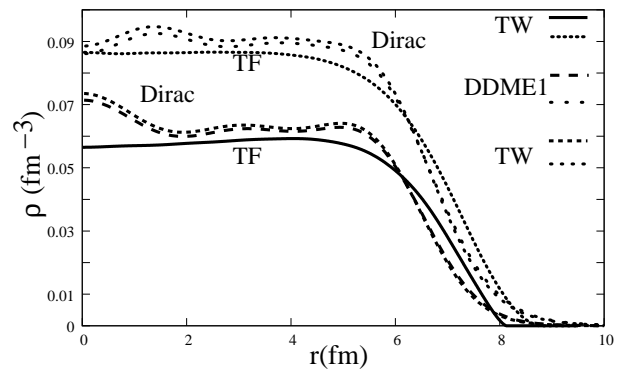}
\end{center}
\caption{$^{208}Pb$ neutron and proton densities obtained with TW
(Thomas-Fermi), DDME1 (Dirac) and TW (Dirac). Larger densities are for
neutrons and lower for protons.}
\label{densities}
\end{figure}

We first plot the {\bf $^{208}Pb$} neutron and proton densities
obtained for the DDH model within the Thomas-Fermi approximation
from \cite{peles} and the solution of the Dirac
equation in Fig. \ref{densities}. As it is well known, the Thomas-Fermi
approximation flattens the central energy densities while the Dirac equation
keeps the internal nuclear structure. The difference in the central density is
then compensated near the surface, with consequences in the calculations of
the mean square radii and neutron skin thickness. One can see that the DDME1
parametrization gives almost the same density profiles as compared with the
original TW parametrization.

In table I we show the results of the present calculations for $^{208}$Pb
obtained with the NL3 and both parametrizations of the DDH model. For the sake
of completeness we also include the results obtained
in \cite{peles} within the Thomas-Fermi approach. One can observe that
the Thomas-Fermi results for the radii are systematically larger than the Dirac
solutions, but the neutron skins are smaller. The surface energies also give
smaller contributions within the TF approach. Both parametrizations of the DDH
model show the same neutron skins albeit the small
differences in the proton and neutron radii. The binding energy and the
surface energy are also slightly different. Finally, the charge radius ($R_c$),
calculated with the charge distribution given by equation (\ref{rhoc}) is
presented and compared with the experimental value. All results are very
similar.

In table II we show the results for $^{40}$Ca, $^{48}$Ca,$^{66}Ni$ and $^{90}$Zr.
Most of the conclusions drawn from Table \ref{tab1} remain valid, i.e., the
neutron skin thickness are the same for both parametrizations of the DDH model,
TW gives larger surface energies than DDME1 and NL3 and the proton, neutron and
charge radii vary from one model to the other.

\begin{figure}
\begin{center}
\includegraphics[width=8.cm]{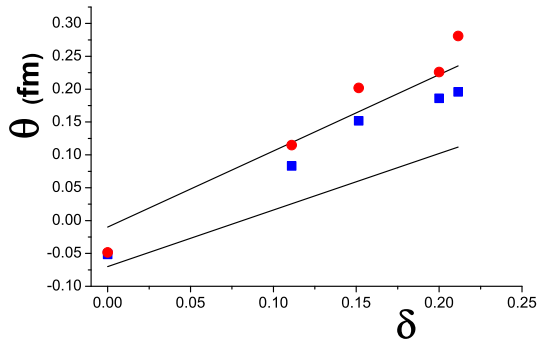} \\
\end{center}
\caption{Skin thickness as a function of the proton-neutron number asymmetry.
Squares represent our values in the TW and full circles in the NL3. The lines
represent the limits for the fitting given in \cite{trzcinska}.
From lower to higher asymmetries we have plotted the results for
$^{40}$Ca, $^{90}$Zr, $^{66}$Ni, $^{48}$Ca and $^{208}$Pb.}
\label{correlation}
\end{figure}

According to \cite{trzcinska}, experimental data show a linear relation
between the neutron skin thickness and the proton-neutron asymmetry of the considered nuclei,
i.e., $\delta=(N-Z)/A$. Assuming that this dependence really exists, the
authors of \cite{trzcinska} fit it as
\begin{equation}
\theta=(-0.04 \pm -0.03) + (1.01 \pm 0.15) \delta.
\end{equation}
In Fig. \ref{correlation} we have plotted the fitted dependence lines
considering the most extreme values for $\theta$. If this dependence has
really to be satisfied, the theoretical points should lie within both curves.
The TW and DDME1 points coincide and they are shown to be inside the
appropriate range while the NL3 results for very asymmetric nuclei are outside
the upper boundary.

\begin{figure}
\begin{center}
\begin{tabular}{cc}
\includegraphics[width=8.cm]{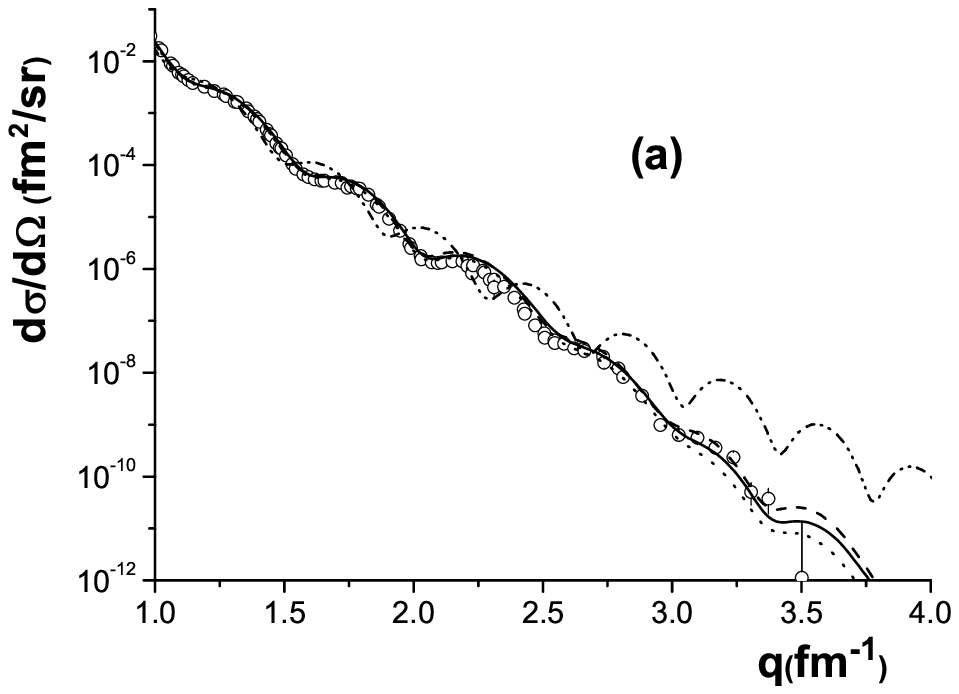}\\
\includegraphics[width=8.cm]{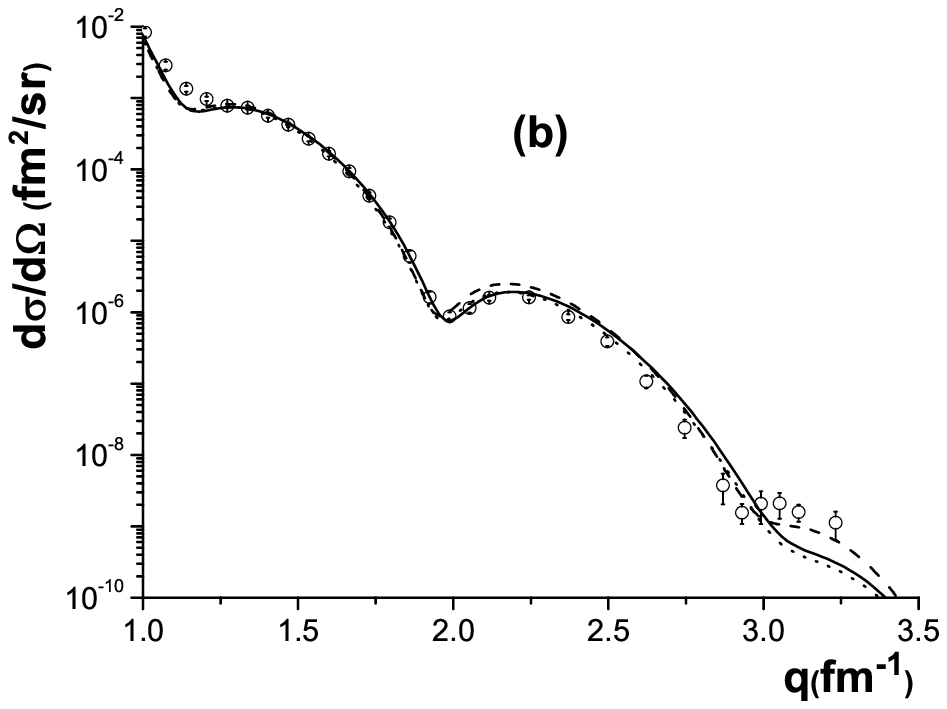} \\
\end{tabular}
\end{center}
\caption{Elastic electron scattering cross section in the NL3 (dashed line),
TW (full line) and DDME1 (dotted line) approaches for a) $^{208}$Pb and 
b) $^{48}$Ca. E is the
electron incident energy taken as 502 MeV for $^{208}$Pb and 
as 757.5 MeV for $^{48}$Ca. Experimental data are from \cite{Frois}
and \cite{belicoso} respectively. The dot-dashed line in a) corresponds to the
Thomas-Fermi approximation and TW parametrization.}
\label{cross}
\end{figure}

Now let's focus on the electron scattering cross sections and the asymmetry. In
Fig. \ref{cross} the elastic cross section for NL3 and DDH models are compared
with the experimental data for $^{208}$Pb and $^{48}Ca$.
Finite size proton as well as
center-of-mass effects were also included in the cross-section calculation.
In both cases the theoretical and experimental results are in very good
agreement and both models show very small quantitative differences along all
the momentum transfer region covered by the data. While TW and DDME1 seem
  to give a slightly better desciption for $^{208}$Pb, NL3 results are closer
  to the cross section for $^{48}Ca$ at momentum transfers between 3.0 and 3.5
fm$^{-1}$. For the sake of completeness we have added the results obtained
with the Thomas-Fermi approximation for the TW parametrization of the DDH
model for $^{208}$Pb.
One can see that the results for the cross section start to deviate
from the experimental points around 1.5 fm$^{-1}$ with larger discrepancies
at larger momentum transfers.

\begin{figure}
\begin{center}
\begin{tabular}{cc}
\includegraphics[width=8.cm]{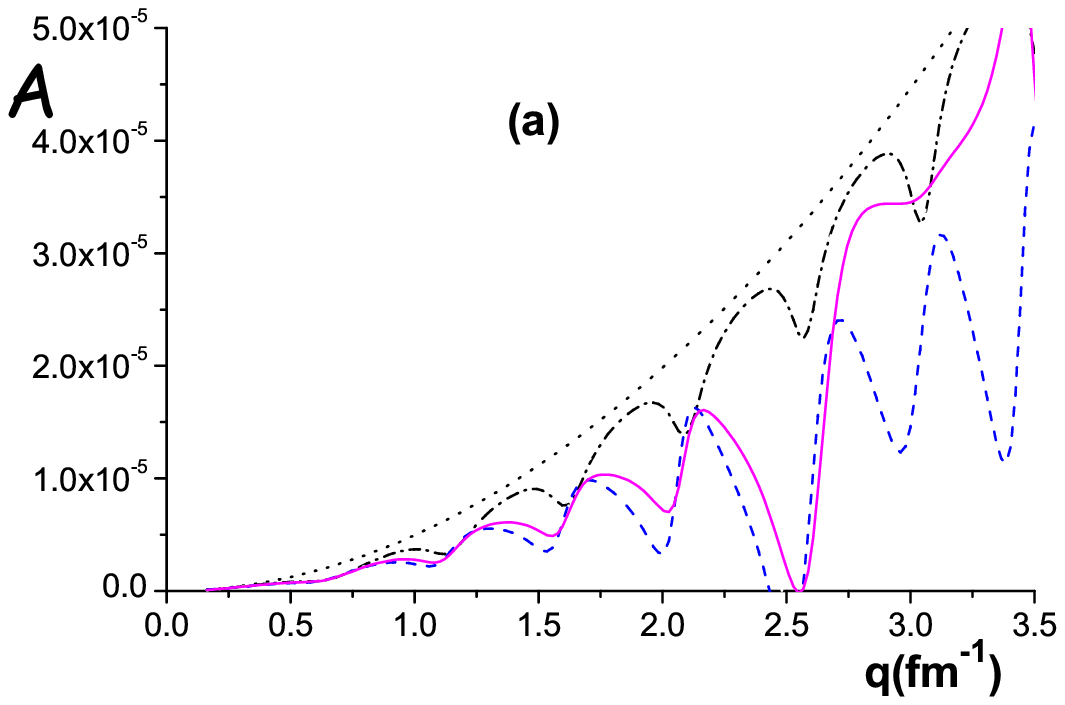} \\
\includegraphics[width=8.cm]{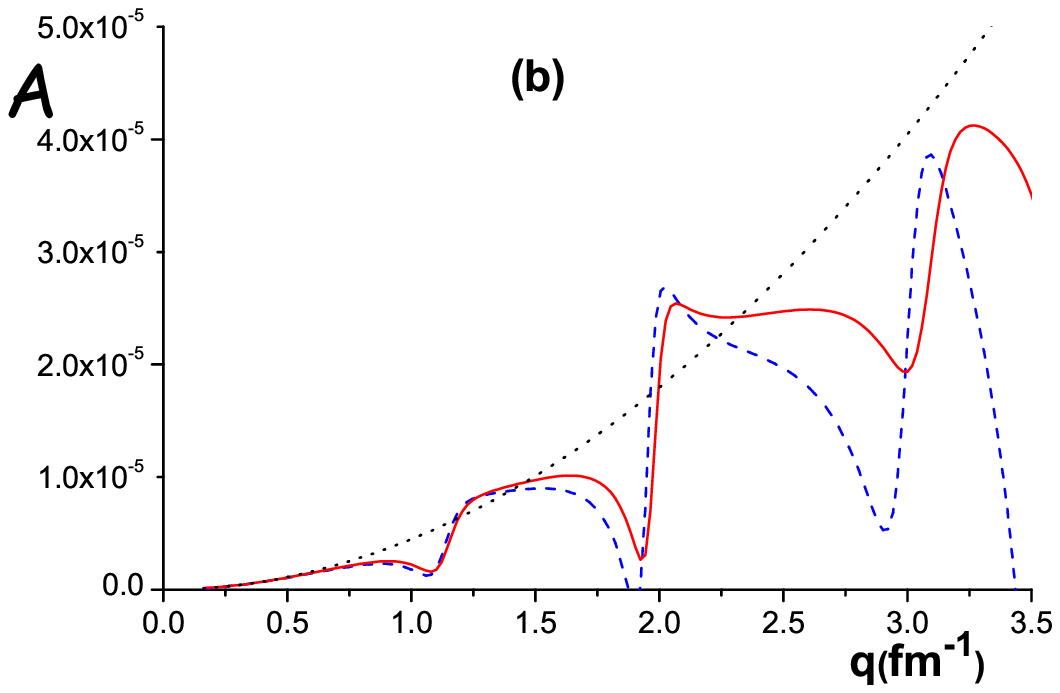}\\
\end{tabular}
\end{center}
\caption{Asymmetry obtained for a) $^{208}$Pb and b) $^{48}Ca$ in the NL3
(dashed line) and TW (full line). The dotted line is the PWBA result and the
  dash-dotted one is the three parameter Fermi result ({\textit{Fermi3p}})
as explained in the text. The incident electron energy was chosen as
$E=800MeV$.}
\label{asym}
\end{figure}

As for the asymmetry, our results are displayed in Fig. \ref{asym} for
$^{208}Pb$.
The curve labeled PWBA is simply the result obtained from equation
(\ref{ass2}) in the Appendix, for the case where
$Z~\rho_n=N~\rho_p$. This can be compared with the curve labeled
{\textit{Fermi3p}}, where the same condition is imposed but
the cross-section calculation considered the partial wave expansion method
and a three parameter Fermi distribution to obtain the Coulomb and weak
potentials as defined in the Appendix. The difference between those curves
clearly states the necessity to incorporate the electron wave distortion
effects in the calculation of the asymmetry. The other two curves give us the
structure effects coming from the NL3 and DDH models respectively. Similar
results are shown in Fig. \ref{asym} for $^{48}Ca$. One should notice that
while the proton distribution in the NL3 and DDH models follow the same trend
even for large momentum transfer, as can be seen from
Fig. \ref {cross}, the asymmetry starts to present important qualitative
differences at $q\gtrsim 2fm^{-1}$, indicating that the neutron distribution
seems to be more sensitive to the model used.

\section{Conclusions}

In the present work we have recalculated the neutron skin thickness and the
asymmetry that will be measured at the PREX experiment with two kinds of
relativistic models previously used within a Thomas-Fermi plus PWBA
approximations. We have here considered the full solution of the Dirac
equation and used an exact calculation for the scattered electron
wavefunction. Density dependent and constant coupling relativistic models
provide different results and so the model dependence of the electron
scattering asymmetry is confirmed by the more exact calculation, although it
is still very hard to be extracted from small momentum transfer data.

If a linear relation between the neutron skin thickness and the proton-neutron
asymmetry of
the considered nuclei is really to be satisfied, as suggested in
\cite{trzcinska}, the TW and DDME1 model parametrizations provide results
within the appropriate range while the NL3 results for very asymmetric nuclei
are outside the upper limit imposed by the present data.

Once a more precise measurement of the neutron skin is published, it will
certainly provide a better constraint to different models and
our results should then be revisited for a proper comparison.

\section{Appendix - Cross Section for Polarized Electrons}

The electron hadron Lagrangian density is
\begin{eqnarray}
{\cal L}=\bar \psi\left[\gamma_\mu\left(i\partial^{\mu}+ e A^\mu
+\gamma_{5}A^{\mu}_{W} \right) -m_{e} \right]\psi + {\cal L}_H,
\end{eqnarray}
\noindent where ${\cal L}_H$ is the hadronic Lagrangian,
$A_{\mu}$ is the electromagnetic field generated by the hadronic target,
\begin{eqnarray}
A^{\mu}_{W}=-\frac{G}{\sqrt{2}}J^{\mu,NC},
\end{eqnarray}
\noindent is the weak field and $G$ the Fermi constant. In the elastic
scattering of electrons from an even-even spherical nucleus whose
{\it elementary} particles are just the protons and neutrons, only the static
time component
of the currents contribute and the electron obeys the Dirac
equation:

\begin{eqnarray}
[\overrightarrow{\alpha}.\overrightarrow{p}+\gamma_{0}m_{e}+V(r)+\gamma_{5}V_{A}(r)]\psi=E\psi,\label{diracI}
\end{eqnarray}
\noindent with $V(r)=-eA_{0}(r)$ and
$V_{A}(r)=\frac{G}{\sqrt{2}}J_{0}^{NC}(r)$. From the standard model we know that the weak neutral
(NC) current is \cite{GreinerBook}:

\begin{eqnarray}
J^{0,NC}(r)=\chi_{p}\rho_{p}(r)+\chi_{n}\rho_{n}(r).
\end{eqnarray}

The above constants are $\chi_{p}\simeq 0.04$ and $\chi_{n}=-0.5$. Hence,
we conclude that the weak potential depends strongly on the
neutron distribution in nuclei, while the e.m. field $A_{0}$ depends
(not considering the neutron electric form factor) exclusively on the proton
distribution.
For high energy electrons and scattering angles not to close to $180^{\circ}$
we may use
 the approximation $\frac{m_{e}}{E}\sim0$, for which the Dirac equation can be
rewritten in the form:

\begin{eqnarray}
[\overrightarrow{\alpha}.\overrightarrow{p}+V(r)\pm
V_{A}(r)]\psi_{\pm}=E\psi_{\pm},\label{diracII}
\end{eqnarray}

\noindent with $\psi_{\pm}=\frac{1}{2}(1\pm\gamma_{5})\psi$ and the $\pm$ signs represent the two possible electron
initial polarization (helicity) states. We have solved the above equation for the electron exactly,
using the well-known partial wave phase-shift expansion method, as explained,
for instance in \cite{Prewitt}.
Note that for each electron helicity state, a different set of phase-shifts must be determined, whose differences
come from the contribution of the weak potential. We now define the asymmetry
through the expression:

 \begin{equation}
\mathcal{A}=\frac{d\sigma_{+}/d\Omega-d\sigma_{-}/d\Omega}{d\sigma_{+}/d\Omega+d\sigma_{-}/d\Omega},\label{ass1}
\end{equation}
where $d\sigma_{\pm}/d\Omega$ is the differential cross section
for initially polarized electrons with positive($+$) and negative
($-$) helicities. The solution of equation (\ref{diracII}) in first order
perturbation theory lead us to the Plane Wave Born Approximation \cite{Don} result:

\begin{equation}
\mathcal{A}=-\frac{Gq^2}{2\pi\alpha\sqrt{2}}[\chi_{p}+\chi_{n}\frac{\rho_{n}(q)}{\rho_{p}(q)}],\label{ass2}
\end{equation}

\noindent where $q$ is the momentum transfer and $\rho_{n(p)}(q)$
being the neutron (proton) distribution in $q$ space. Even for medium mass
nuclei, important differences between the PWBA and the exact result should be
found in the asymmetry \cite{horo98}. In this paper, all the cross sections
were obtained through
the full solution of the Dirac equation for the electron.

\section*{ACKNOWLEDGMENTS}

This work was partially supported by CNPq(Brazil).

\newpage

\begin{table}
\caption{$^{208}$ Pb properties}
\label{tab1}
\begin{center}
\begin{tabular}{ccccccccccc}
\hline
model & approximation & $R_n$ & $R_p$ & $R_c$ & $\theta$ & $B/A$ & $\sigma$ \\
&& (fm) & (fm) & (fm) & (fm) & MeV & Mev/fm$^2$ \\
\hline
NL3 & TF & 5.79 & 5.57 &      & 0.22 & -7.79 & 0.96\\
NL3 & Dirac & 5.74 & 5.46 & 5.51 & 0.28 & -7.91 & 1.13\\
\hline
TW & TF & 5.68 & 5.52 &      & 0.16 & -7.46 & 1.10\\
TW & Dirac & 5.61 & 5.42 & 5.48 & 0.20 & -7.78 & 1.30\\
\hline
DDME1 & Dirac & 5.66 & 5.46 & 5.51 & 0.20 &-7.91 & 1.18\\
\hline
exp.\cite{fricke} & & & & 5.50 &  & & \\
exp. \cite{audi}  & & & &      &                & -7.87 & \\
exp. \cite{kraszna}& & & &     & $0.12\pm 0.07$  &  & \\
exp. \cite{hintz} & & &  &     & $0.20\pm 0.04$  &  & \\
exp. \cite{klos} & & & & & $0.16 \pm 0.02 \pm 0.04$ & & \\

\hline
\end{tabular}
\end{center}
\end{table}


\begin{table}[h]
\caption{Finite nuclei properties}
\label{tab2}
\begin{center}
\begin{tabular}{ccccccccccc}
\hline
model & nuclei & $R_n$ & $R_p$ & $R_c$ & $\theta$ & $B/A$ & $\sigma$ \\
&& (fm) & (fm) & (fm) & (fm) & MeV & Mev/fm$^2$ \\
\hline
NL3 & $^{40}$Ca & 3.32 & 3.37 & 3.43 &-0.05 & -8.62 & 1.48 \\
TW & $^{40}$Ca & 3.28 & 3.33 & 3.39 &-0.05 & -8.36 & 1.60  \\
DDME1 & $^{40}$Ca & 3.32 & 3.37 & 3.43 &-0.05 & -8.62 & 1.45  \\
exp.\cite{twring}  & $^{40}$Ca & & & 3.48  &  & -8.55 & \\
\hline
NL3 & $^{48}$Ca & 3.60 & 3.37 & 3.44 &0.23 & -8.72 & 1.54 \\
TW &  $^{48}$Ca & 3.54 & 3.35 & 3.42 &0.19 & -8.49 & 1.70  \\
DDME1 & $^{48}$Ca & 3.58 & 3.39 & 3.46 &0.19 & -8.66 & 1.53  \\
exp.\cite{web}  &  $^{48}$Ca & &  & & & -8.67 & \\
exp.\cite{twring}  &  $^{48}$Ca & & & 3.48  &  &  & \\
\hline
NL3 &  $^{90}$Zr & 4.30 & 4.19 & 4.25 &0.11 & -8.86 & 1.37 \\
TW &  $^{90}$Zr & 4.24 & 4.15 & 4.22 &0.08 & -8.55 & 1.52 \\
DDME1 & $^{90}$Zr & 4.28 & 4.19 & 4.25 &0.08 & -8.73 & 1.38 \\
exp.\cite{web} & $^{90}$Zr &  &  &  & & -8.71 &  \\
\hline
NL3 &  $^{66}$Ni & 3.96 & 3.76 & 3.82 &0.20 & -8.74 & 1.47 \\
TW &  $^{66}$Ni & 3.89 & 3.74 & 3.81 &0.15 & -8.56 & 1.63 \\
DDME1 & $^{66}$Ni & 3.93 & 3.77 & 3.84 &0.16 & -8.72 & 1.49 \\
exp.\cite{web} & $^{66}$Ni &  &  &  & & -8.74 &  \\
\hline
\end{tabular}
\end{center}
\end{table}


\begin{thebibliography}{99}

\bibitem{sw} B. Serot and J.D. Walecka, {\em Advances in Nuclear
Physics} 16, Plenum-Press, (1986) 1.

\bibitem{original} H. Lenske and C. Fuchs, Phys. Lett. {\bf B 345}, 355
  (1995); C. Fuchs, H. Lenske and H.H. Wolter, Phys. Rev. {\bf C 52}, 3043
(1995).

\bibitem{tw} S. Typel and H. H. Wolter, Nucl. Phys. {\bf A656}, 331 (1999).

\bibitem{gaitanos} T. Gaitanos, M. Di Toro, S. Typel, V. Baran, C. Fuchs,
V. Greco  and H. H. Wolter, Nucl. Phys. {\bf A732}, 24 (2004).

\bibitem{br} G.E. Brown and M. Rho, Phys. Rev. Lett. {\bf 66}, 2720 (1991).

\bibitem{nl3} G. A. Lalazissis, J. K\"onig and P. Ring,
Phys. Rev. C {\bf 55}, 540 (1997).

\bibitem{tm1} K. Sumiyoshi, H. Kuwabara, H. Toki, Nucl. Phys. {\bf A
581}, 725 (1995).

\bibitem{glen} N. K. Glendenning, Compact Stars, Springer-Verlag, New-York,
2000.

\bibitem{nlwr} C.J. Horowitz and J.Piekarewicz, Phys. Rev.{\bf C 64},
 062802R (2001); J.K. Bunta and S. Gmuca, Phys. Rev. C {\bf 68}, 054318 (2003);
J.K. Bunta and S. Gmuca, Phys. Rev. C {\bf 70}, 054309 (2004).

\bibitem{vries} H. de Vries, C.W. de Jager and C. de Vries, Atomic and Nuclear
Data Tables {\bf 36}, 495 (1987).

\bibitem{horo} C.J. Horowitz, S.J. Pollock, P.A. Souder and R. Michaels,
  Phys. Rev. {\bf C 63}, 025501 (2001).

\bibitem{Don} T.W. Donnelly, J. Dubach and I. Sick, Nucl. Phys.
{\bf A503 } 589 (1989).

\bibitem{prex} K.A. Aniol et al. (HAPPEX) (2005), nucl-ex/0506010; {\it
  ibidem}, nucl-ex/0506011; R. Michaels, P.A. Souder and G.M. Urciuoli (2005),
  URL http://hallaweb.jlab.org/parity/prex.

\bibitem{trzcinska} A. Trzcinska, J. Jastrzebbski, P. Lubinski, F.J. Hartmann,
  R. Schmidt, T. von Egidy and B. Klos, Phys. Rev. Lett. {\bf 87}, 082501
  (2001).

\bibitem{klos} B. Klos et al., nucl-ex/0702016.

\bibitem{peles} S.S. Avancini, J.R. Marinelli, D. P. Menezes, M.M. W. Moraes
and C. Provid\^encia, {\it Phys. Rev.} {\bf C 75} (2007) 055805.

\bibitem{horo98} C.J. Horowitz, Phys. Rev. {\bf C57}, 3430 (1998).

\bibitem{twring} T. Niksic, D. Vretenar, P. Finelli and P. Ring,
  Phys. Rev. {\bf C 66}, 024306 (2002); D. Vretenar, T. Niksic and P. Ring,
  Phys. Rev. {\bf C 68}, 024310 (2003).

\bibitem{ring} Y.K. Gambhir, P. Ring and A. Thimet, Ann. Phys. {\bf 198}, 132
  (1990).

\bibitem{ddpeos} S.S. Avancini and D.P. Menezes, Phys. Rev. {\bf C 74},
015201 (2006).

\bibitem{negele} J.W. Negele, Phys. Rev. {\bf C 1}, 1260 (1970).

\bibitem{Frois} B. Frois et al., Phys. Rev. Letters {\bf 38 }, 152 (1977).

\bibitem{belicoso} J.B. Bellicard et al, Phys. Rev. {\bf C21}, 1652
(1980).

\bibitem{fricke} G.Fricke, C. Bernhardt, K.Heilig, L.A. Schaller, L.
Schellinberg, E.B. Shera,C.W. de Jager, At. Data Nucl. Data Tables {\bf 60}
(1995)177.

\bibitem{audi} G. Audi, A.H. Waptra, C. Thibault, Nucl. Phys. {\bf A 729},
337 (2003).

\bibitem{kraszna}A. Krasznahorkay et a., Nucl. Phys. {\bf A 731}, 224 (2004).

\bibitem{hintz}V.E. Starodubsky, N.M. Hintz,Phys. Rev. {\bf C49},2118(1994).


\bibitem{GreinerBook} W. Greiner and B. Muller, Gauge Theory of the Weak
Interactions (Springer, 1996).

\bibitem{Prewitt} J. F. Prewitt and L. W. Wright, Phys. Rev. {\bf C9}, 2033
(1974).

\bibitem{web} Data obtained from www.nndc.bnl.gov; visited on 11 July 2007.

\end{thebibliography}
\end{document}